\documentclass[a4paper,12pt]{article}
\usepackage[utf8]{inputenc}
\usepackage{graphicx}
\usepackage{amssymb}
\usepackage{bm}
\usepackage{lineno}

\begin{document}

\begin{center}
\begin{Large} \textbf{Bottomonium production at forward rapidity with ALICE at
the LHC}
\end{Large} \\
\vspace{0.75cm}
\begin{large}
Massimiliano Marchisone, for the ALICE Collaboration
\end{large}
\vspace{0.5cm}
\begin{footnotesize}
Dipartimento di Fisica Sperimentale dell'Universit\`{a} di Torino\\and
Sezione INFN di Torino, Turin, Italy
\end{footnotesize} \\
\vspace{0.5cm}
Presented at the Initial Stages 2014 Conference, Napa, 3--7 December 2014
\end{center}
\vspace{0.5cm}
\begin{abstract}
\begin{footnotesize}
Bottomonium production is a powerful tool to investigate hadron collisions and
the properties of the medium created in heavy-ion collisions. According to the
color-screening model, these mesons give important information about the
deconfined medium called Quark-Gluon Plasma (QGP) produced in ultrarelativistic
heavy-ion collisions. Cold nuclear matter (CNM) effects can modify the
bottomonium production even in absence of deconfined matter: the study of
proton-nucleus collisions is therefore essential to disentangle these effects
from the hot ones. Last but not least, measurement in pp collisions serve as
crucial test of different QCD models of quarkonium hadroproduction and provide
the reference for the study in nucleus-nucleus collisions.

In ALICE, bottomonium is measured at forward rapidity ($2.5 < y < 4$) down to
zero transverse momentum, exploiting the dimuon decay channel. The latest
results in pp, Pb--Pb and p--Pb collisions are discussed and
compared to theoretical calculations.
\end{footnotesize}
\end{abstract}

\begin{small}

\section{Introduction}

According to the color-screening model \cite{screening}, the dissociation
probability of the different quarkonium states ($c\bar{c}$ and $b\bar{b}$
mesons) due to the Quark-Gluon Plasma (QGP) is expected to provide essential
information about the properties of the system produced in heavy-ion collisions
(AA). Competing mechanisms called cold nuclear matter (CNM) effects (such as
gluon shadowing or coherent parton energy loss) can modify the quarkonium
production even in absence of the QGP \cite{CNM}, thus complicating the
interpretation of the results. Data from proton-nucleus collisions (pA) are
therefore necessary to disentangle these effects from the hot ones. Finally, new
measurements in pp collisions help to constrain the various models describing
the quarkonium production mechanisms.

The study of bottomonium production also complements the results obtained with
charmonia: for the latter system an important regeneration in AA collisions
might be expected at the LHC energies due to the large number of $c\bar{c}$ pair
produced, while this effect should be much smaller for the bottomonium
\cite{regeneration}. Moreover, the measurement of $\Upsilon$ allows a study in a
different Bjorken-$x$ range with respect to the J/$\psi$ and the theoretical
calculations for bottomonium are more robust due to the higher mass of $b$
quark.

ALICE \cite{ALICE} is the LHC experiment dedicated to the study of heavy-ion
collisions and has collected data in pp, p--Pb and Pb--Pb collisions.
At forward rapidity ($2.5<y<4$) quarkonia are reconstructed with the muon
spectrometer down to a transverse momentum
($p_\mathrm{T}$) equal to zero, exploiting their decay into $\mu^+\mu^-$.

\section{$\Upsilon$ production in pp collisions}

The $\Upsilon(1S)$ and $\Upsilon(2S)$ production cross sections have been
measured at forward rapidity in pp collisions at $\sqrt{s}=7$ TeV
\cite{ALICE_pp}. The rapidity dependence presented in Fig. \ref{fig: pp_sigma}
(left) shows a good agreement for both resonances with the measurements of
LHCb \cite{LHCb_pp} in the same range and complements the results obtained by
CMS at midrapidity \cite{CMS_pp_1,CMS_pp_2}.

\begin{figure}[htbp]
 \centering
 \includegraphics[width=0.45\textwidth]{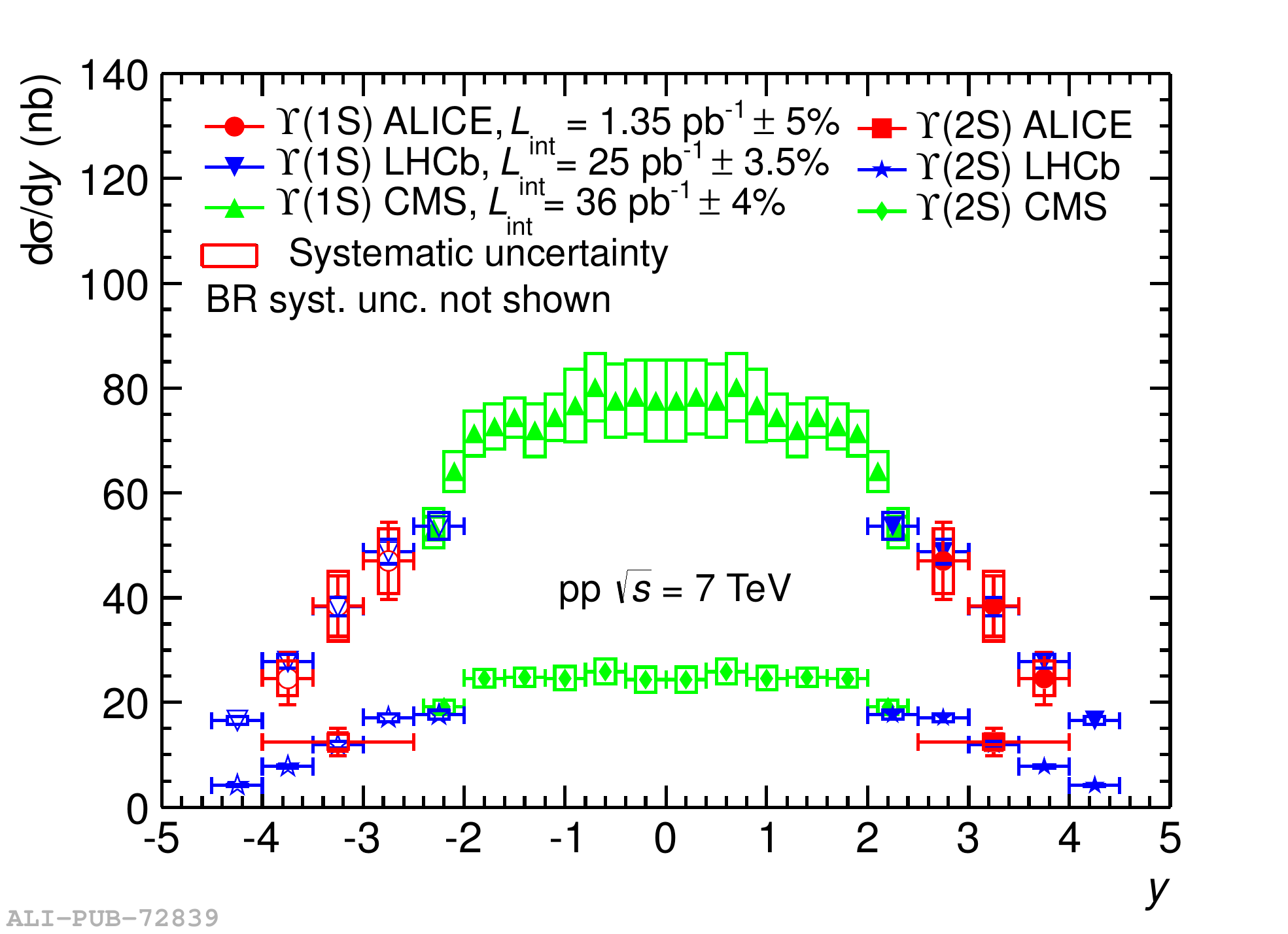}
 \includegraphics[width=0.42\textwidth]{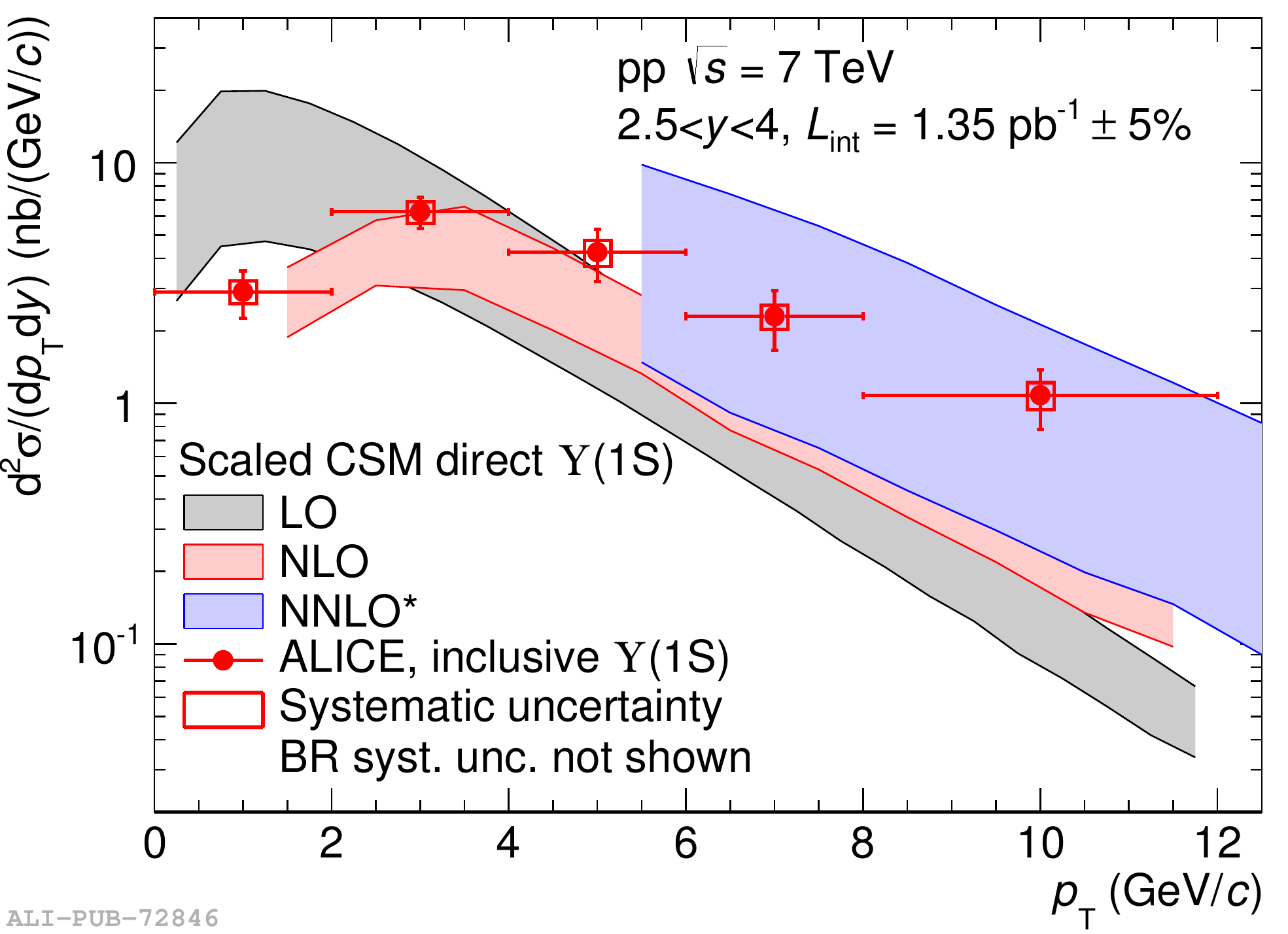}
 \caption{\small On the left, rapidity differential cross sections of
$\Upsilon(1S)$ and $\Upsilon(2S)$ measured by ALICE, LHCb \cite{LHCb_pp} and CMS
\cite{CMS_pp_1,CMS_pp_2}. On the right, $p_{\mathrm{T}}$ differential cross
section of $\Upsilon(1S)$ compared to three theoretical calculations
\cite{pp_th_1}.}
 \label{fig: pp_sigma}
\end{figure}

In the right panel of Fig. \ref{fig: pp_sigma} the inclusive $\Upsilon(1S)$
production cross section as a function of $p_{\mathrm{T}}$ is compared to
Color Singlet Model (CSM) calculations which account for the feed-down from
higher mass states \cite{pp_th_1}. The leading order (LO) calculation
underestimates the data for $p_{\mathrm{T}} > 4$ GeV/c and falls too rapidly
with increasing $p_{\mathrm{T}}$. The $p_{\mathrm{T}}$ dependence of the
next-to-leading order (NLO) calculation is closer to the measurements, but the
prediction still underestimates the cross section over the full $p_{\mathrm{T}}$
range. A good agreement is achieved at a leading-$p_\mathrm{T}$
next-to-next-to-leading order (NNLO*), but over a limited $p_{\mathrm{T}}$ range
and with large theoretical uncertainties.

\section{$\Upsilon$ production in Pb--Pb collisions}

The effects of the hot and dense medium on the $\Upsilon(1S)$ production at
forward rapidity in Pb--Pb collisions at $\sqrt{s_\mathrm{NN}}=2.76$ TeV are
quantified by means of the nuclear modification factor ($R_{\mathrm{AA}}$),
defined as the meson yield in Pb--Pb divided by the production cross section in
pp collisions and the nuclear overlap function \cite{ALICE_PbPb}.

The $R_{\mathrm{AA}}$ in Fig. \ref{fig: Raa_Emerick} (left) shows a more
pronounced $\Upsilon(1S)$ suppression in central than in semiperipheral
collisions. Moreover, the rapidity dependence in the right panel suggests a
stronger suppression at forward than at midrapidity as it appears from the
comparison with the CMS point in $|y|< 2.4$ \cite{PbPb_CMS}.

\begin{figure}[htbp]
 \centering
 \includegraphics[width=0.45\textwidth]{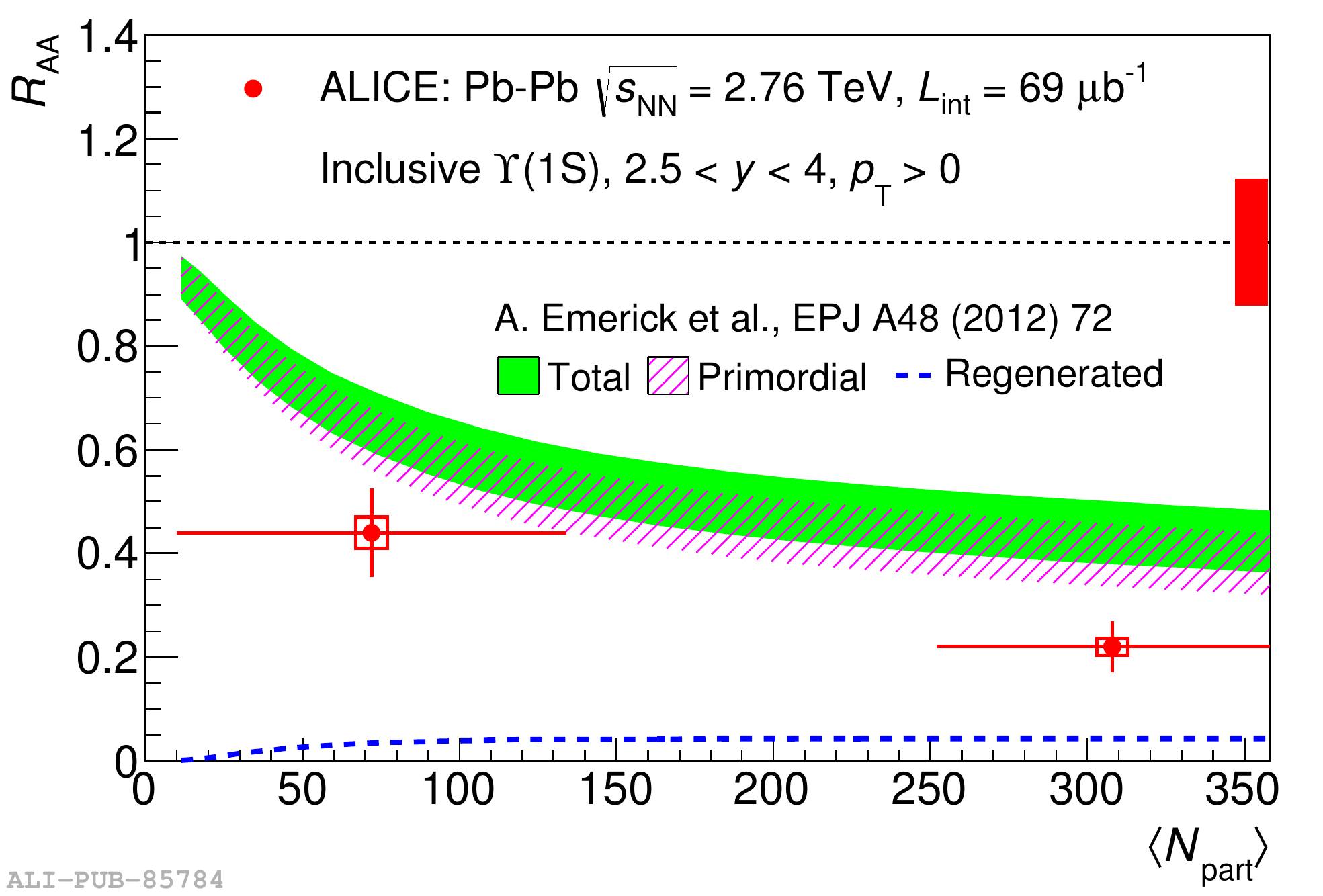}
 \includegraphics[width=0.45\textwidth]{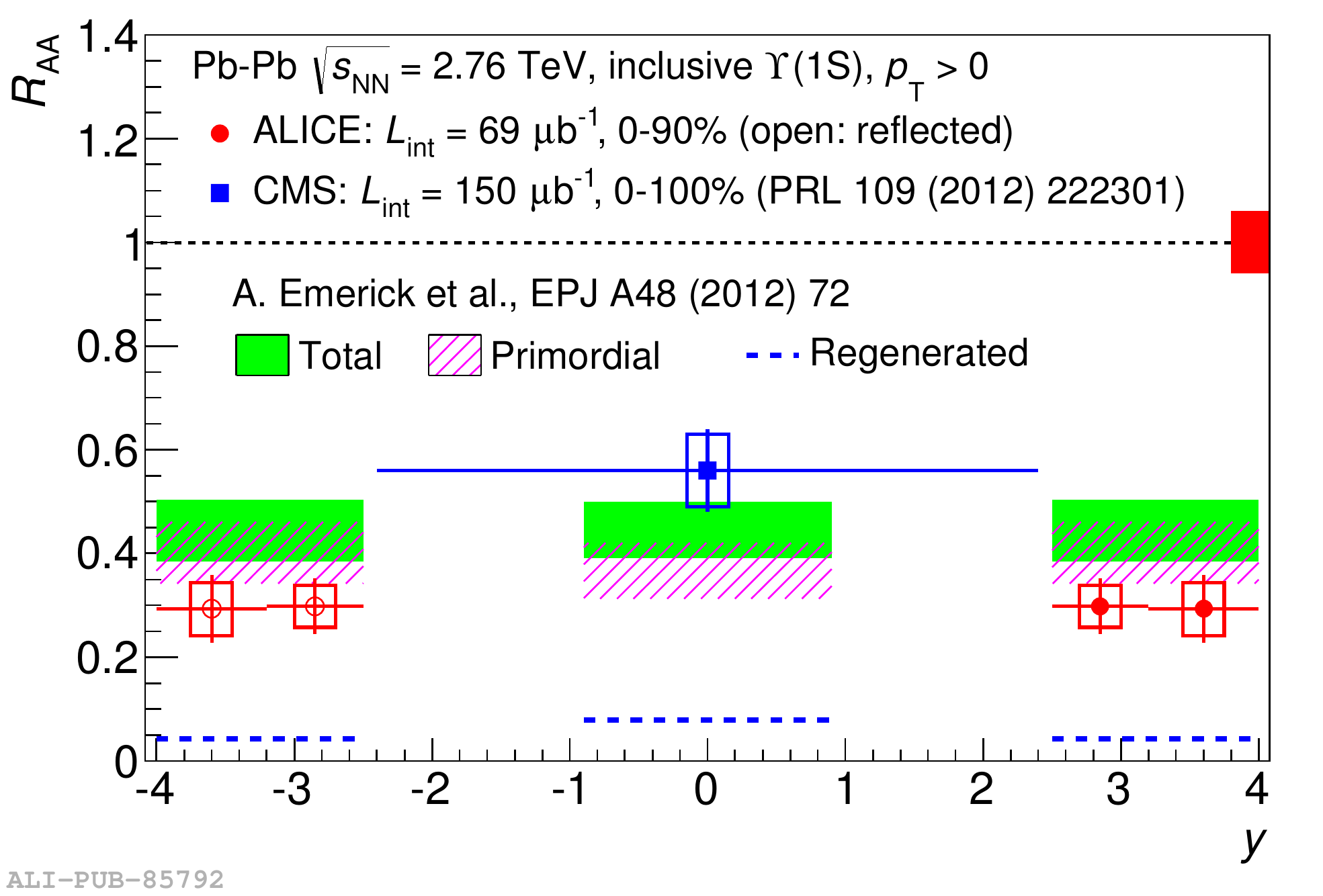}
 \caption{\small Inclusive $\Upsilon(1S)$ $R_{\mathrm{AA}}$ as a function
of the average number of participants (left) and rapidity (right) compared to
predictions from a transport model \cite{upsilon_Emerick}.}
 \label{fig: Raa_Emerick}
\end{figure}

Predictions from a transport model are also shown in the same figures.
The calculation is based on a kinetic rate-equation approach in an evolving
QGP and include both suppression and regeneration effects
\cite{upsilon_Emerick}. CNM effects are calculated by varying an effective
absorption cross section between 0 and 2 mb, resulting in an uncertainty band.
This model underestimates the observed suppression, even if the centrality
dependence is fairly reproduced. The model predicts also an almost constant
$R_{\mathrm{AA}}$ as a function of the rapidity which is in disagreement with
the trend observed by ALICE and CMS.

Other predictions based on a dynamical model \cite{upsilon_Strickland} or
another transport model \cite{upsilon_Zhou}, not described here, show the same
difficulty to reproduce the ALICE data.

\section{$\Upsilon$ production in p--Pb collisions}

The nuclear modification factor ($R_{\mathrm{pPb}}$) measured  in p--Pb
collisions at $\sqrt{s_{\mathrm{NN}}}=5.02$ TeV is used to determine the CNM
effects \cite{ALICE_pPb}. As shown in the two panels of Fig. \ref{fig: RpA}, the
inclusive $\Upsilon(1S)$ production is suppressed at forward rapidity (p-going
direction), while at backward rapidity (Pb-going direction) the measurement is
compatible with unity within uncertainties, disfavoring a strong gluon
antishadowing. At forward rapidity the $\Upsilon(1S)$ and J/$\psi$
$R_\mathrm{pPb}$ are rather similar. At backward rapidity, the J/$\psi$
measurement is systematically above that of $\Upsilon(1S)$, even if they are
still consistent within uncertainties \cite{jpsi_pPb}. Finally the
$R_\mathrm{pPb}$ measured by LHCb \cite{LHCb_pPb} is consistent within
uncertainties with the ALICE result.

\begin{figure}[htbp]
 \centering
 \includegraphics[width=0.45\textwidth]{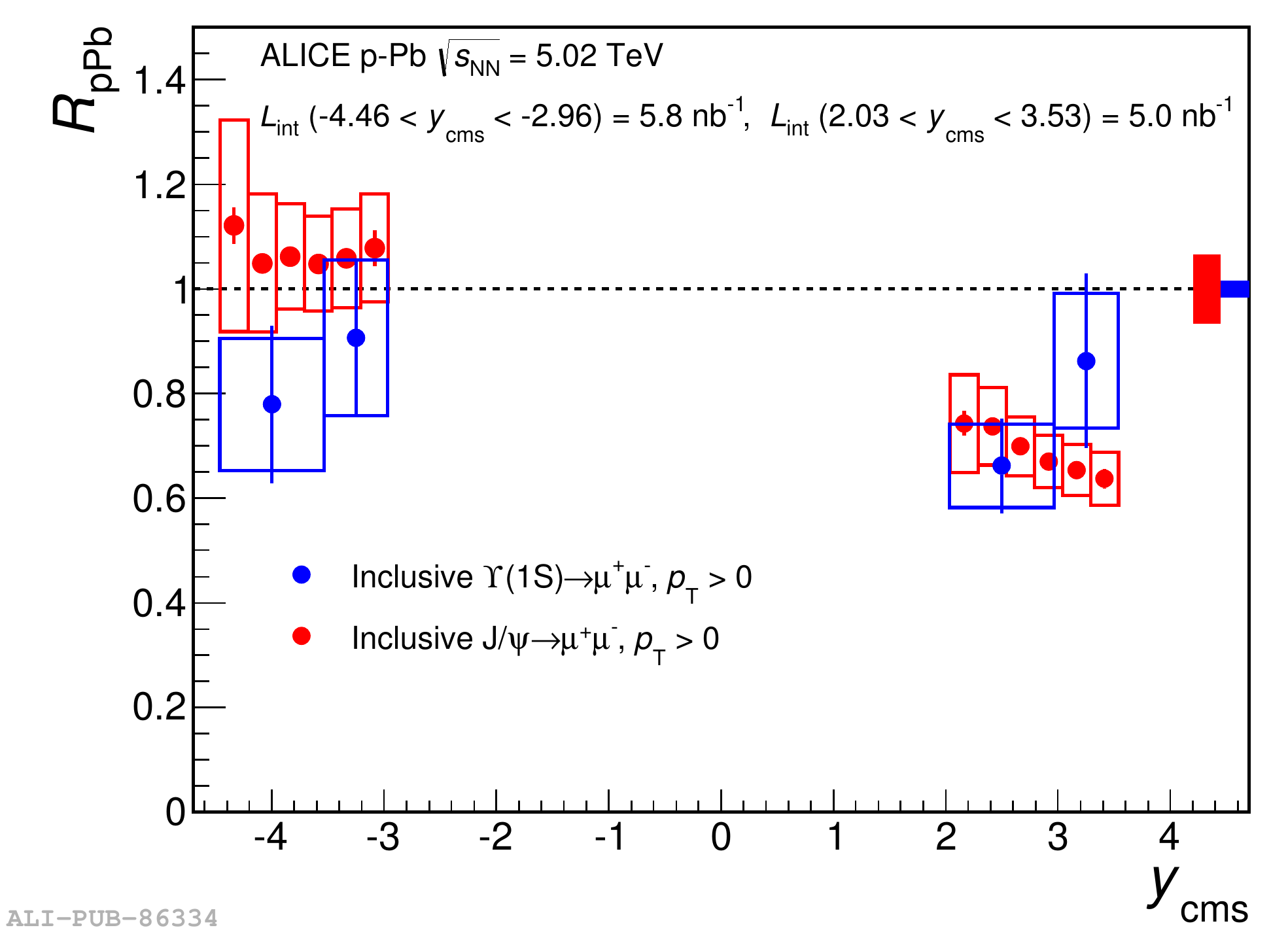}
 \includegraphics[width=0.45\textwidth]{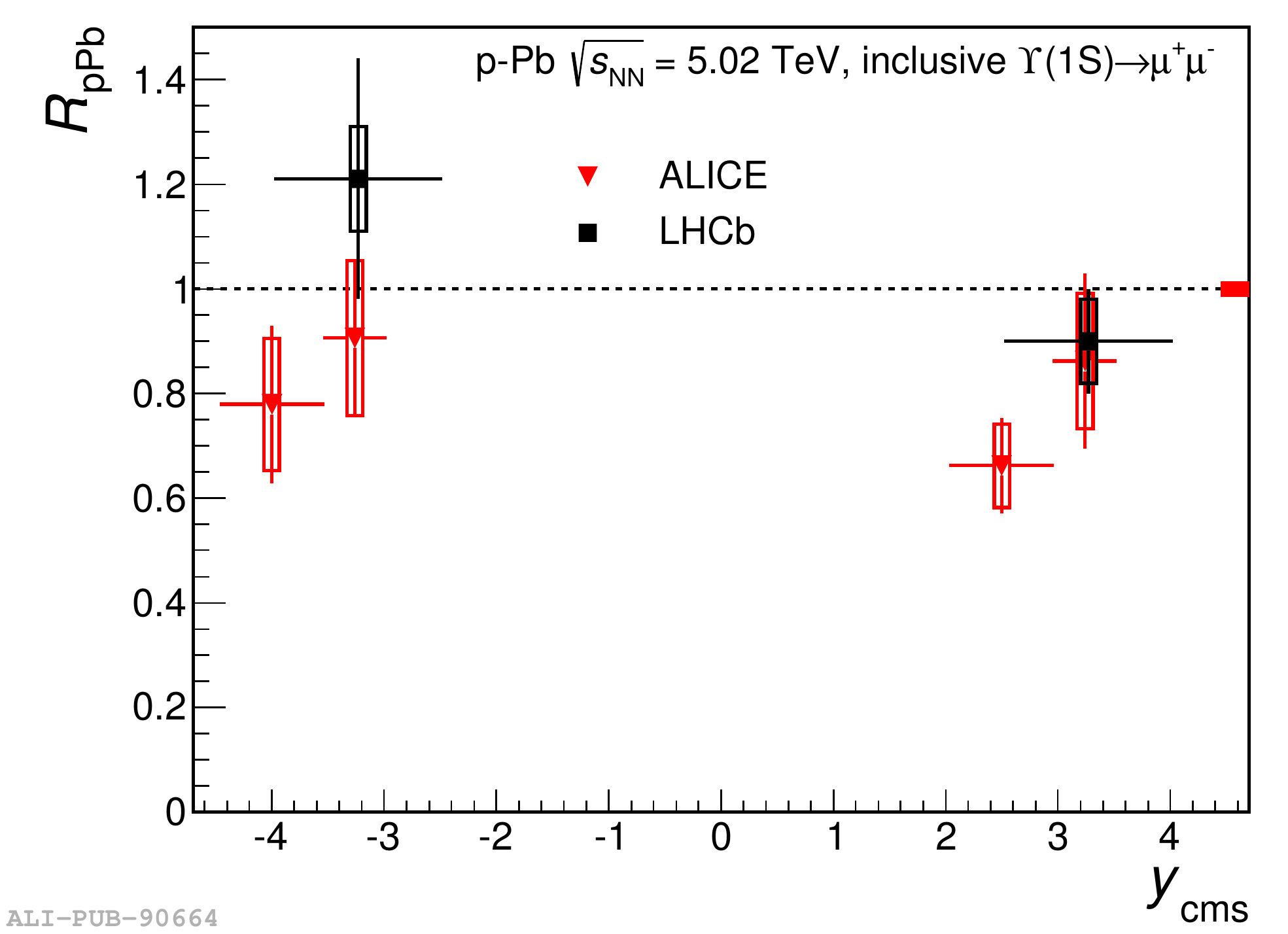}
 \caption{\small $R_{\mathrm{pPb}}$ of inclusive $\Upsilon(1S)$ as a function of
rapidity compared to the analogous measurements for the J/$\psi$ made by ALICE
\cite{jpsi_pPb} (left) and for $\Upsilon(1S)$ made by LHCb \cite{LHCb_pPb}
(right).}
 \label{fig: RpA}
\end{figure}

In the left plot of Fig. \ref{fig: RpA_model}, the ALICE results have been
compared to a NLO Color Evaporation Model (CEM) calculation with shadowing
parametrized by EPS09 at NLO \cite{upsilon_EPS09} which tends to overestimate
the observed $\Upsilon(1S)$ $R_{\mathrm{pPb}}$. Coherent parton energy loss
calculations \cite{upsilon_eloss_1} with or without EPS09 are also shown: the
former reproduces the data at forward rapidity, while the latter is in better
agreement with the measurements at backward rapidity.

\begin{figure}[htbp]
 \centering
 \includegraphics[width=0.45\textwidth]{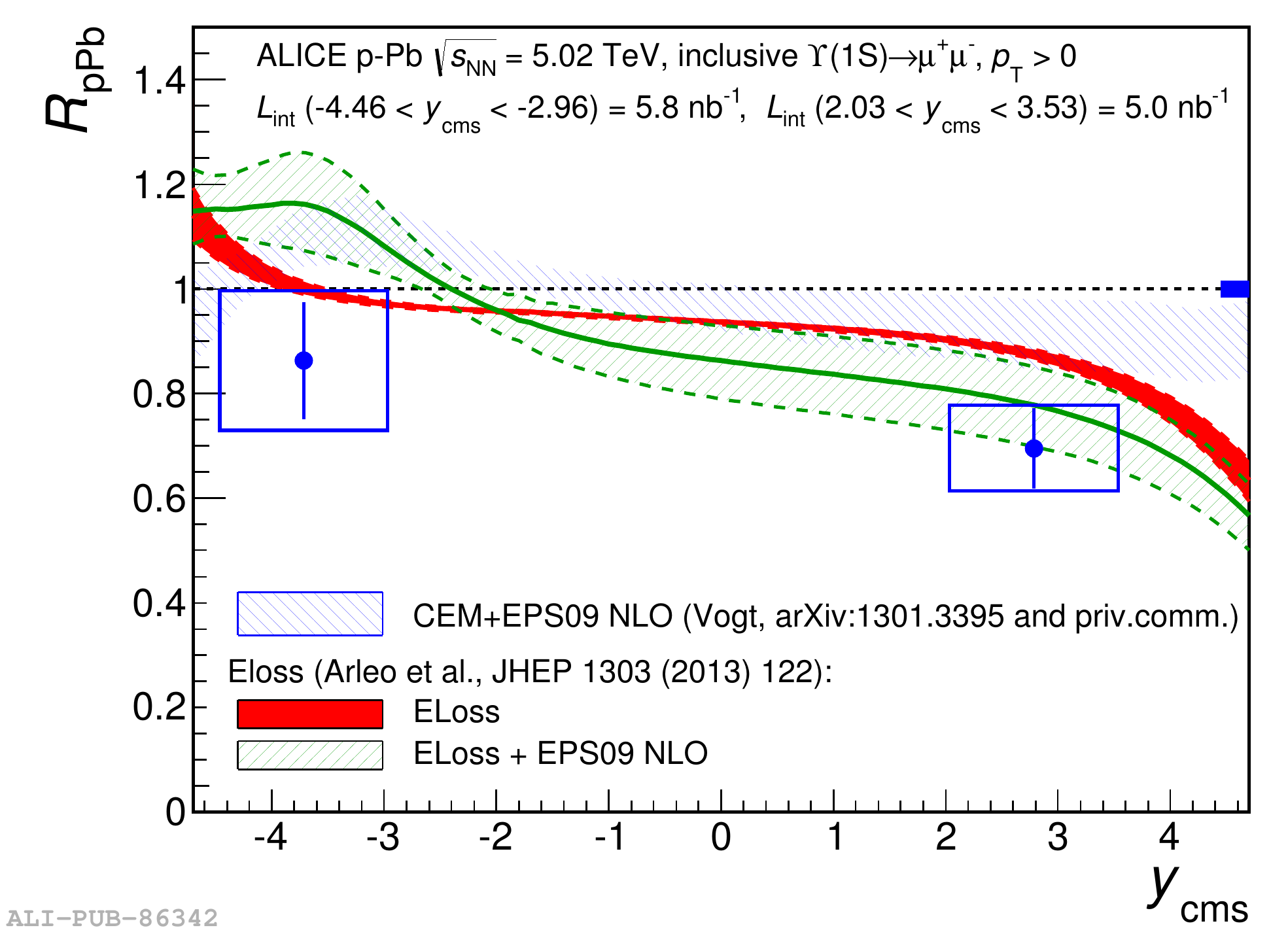}
 \includegraphics[width=0.45\textwidth]{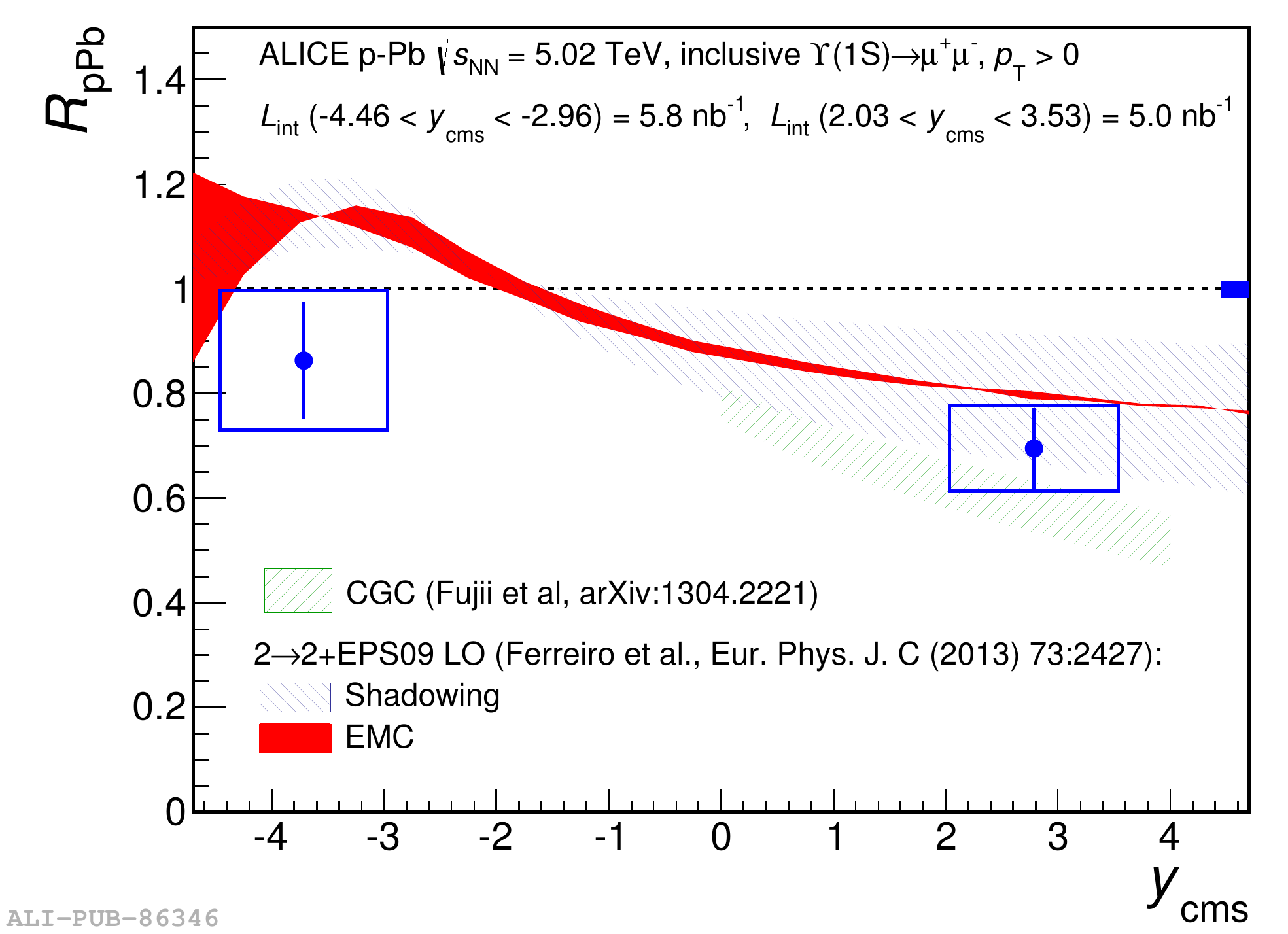}
 \caption{\small Nuclear modification factor of inclusive $\Upsilon(1S)$ in
p--Pb collisions as a function of rapidity compared to several model
calculations
(\cite{upsilon_EPS09,upsilon_eloss_1,upsilon_eloss_2,upsilon_CGC}).}
 \label{fig: RpA_model}
\end{figure}

In the right panel, the results are compared to a LO calculation of a
$gg\rightarrow\Upsilon g$ production with shadowing parametrization
\cite{upsilon_eloss_2}. The two bands show the uncertainties related to EPS09 LO
in the shadowing region and in the EMC region. A calculation at forward
rapidity based  on the CGC framework coupled with a CEM production is also shown
\cite{upsilon_CGC}. Although this prediction only slightly
underestimates the $\Upsilon(1S)$ $R_{\mathrm{pPb}}$, it is not able to
reproduce the analogous J/$\psi$ measurement in the same rapidity range
\cite{jpsi_pPb}.

Finally, the forward-to-backward ratio ($R_\mathrm{FB}$) defined as the ratio
of the nuclear modification factors at forward and backward rapidities is shown
in Fig. \ref{fig: RFB}. It has the advantage to be independent from the pp cross
section which represents the main source of systematic uncertainties, but it can
only be measured in the restricted rapidity range $2.96<|y_\mathrm{cms}|<3.53$.
The $R_\mathrm{FB}$ is compatible with unity and is larger than that of the
J/$\psi$ \cite{jpsi_pPb}. All models describe the data within the present
uncertainties of the measurement.

\begin{figure}[htbp]
 \centering
 \includegraphics[width=0.45\textwidth]{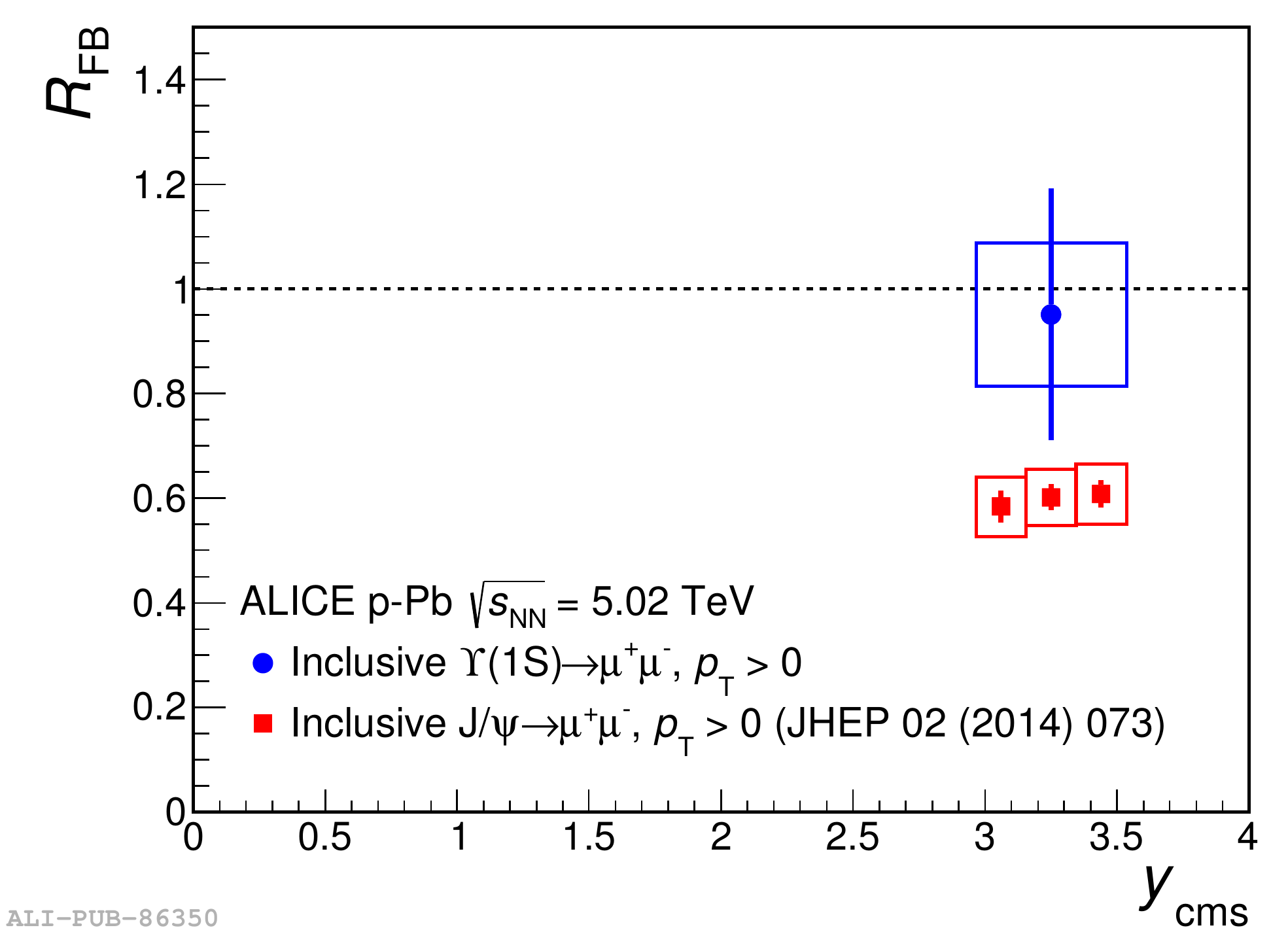}
 \includegraphics[width=0.45\textwidth]{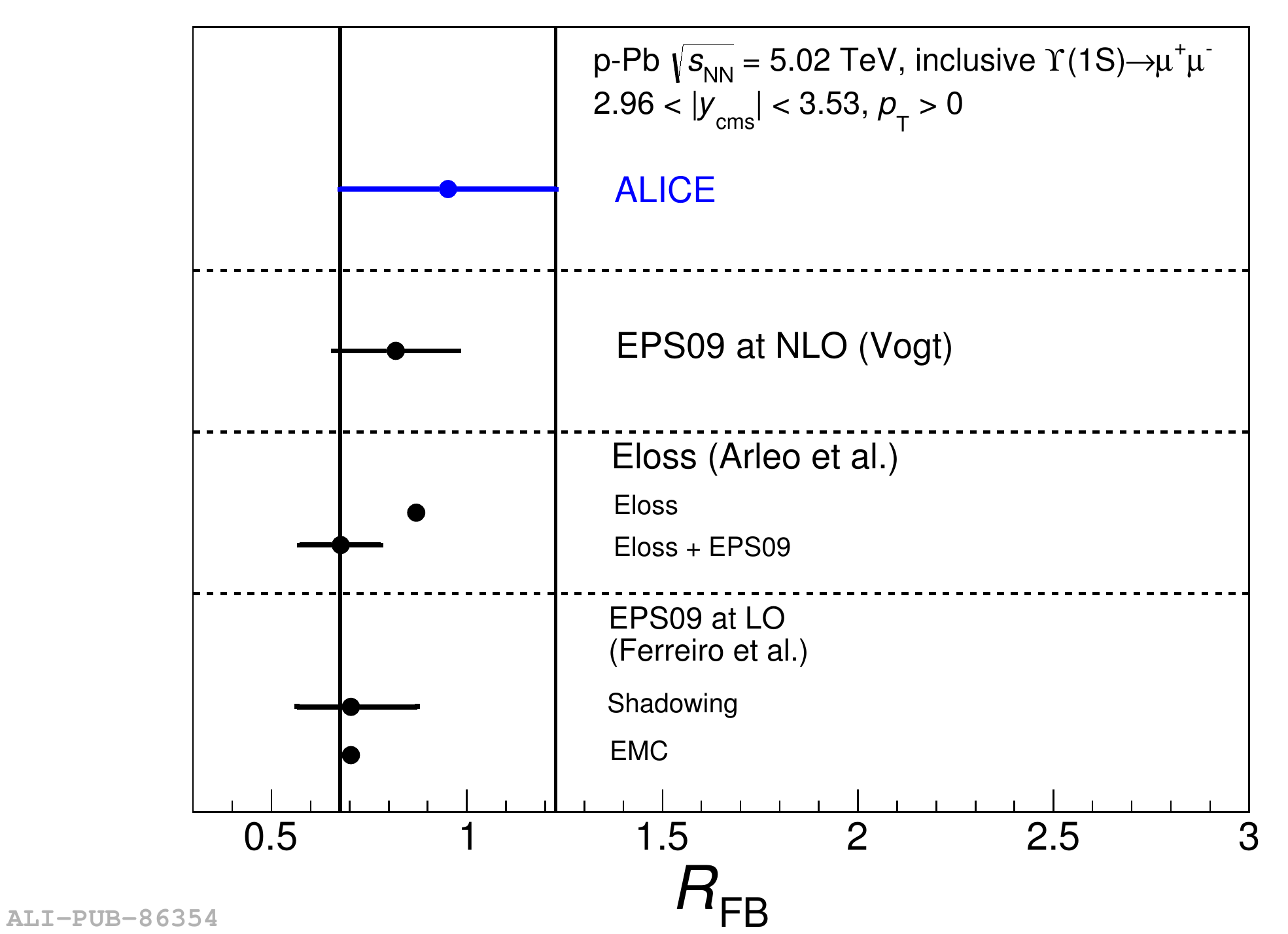}
 \caption{\small Forward-to-backward ratio of inclusive $\Upsilon(1S)$ yields
compared to the J/$\psi$ data \cite{jpsi_pPb} (left) and to the theoretical
models previously described (right).}
 \label{fig: RFB}
\end{figure}

\section{Conclusions}

The measured inclusive production cross sections of $\Upsilon(1S)$ and
$\Upsilon(2S)$ at forward rapidity in pp collisions are in good agreement with
the results obtained by LHCb and complement those by CMS at midrapidity.
Both CSM LO and NLO calculations underestimate the data at large transverse
momentum, while the addition of the NNLO* contributions
helps to reduce this disagreement, but with large theoretical uncertainties.

The observed suppression of inclusive $\Upsilon(1S)$ in Pb--Pb collisions
increases with the centrality and rapidity as shown in the large domain covered
by ALICE and CMS. The suppression, larger than what predicted by the models
considered, might point to a significant dissociation of direct $\Upsilon(1S)$.

Finally, the $\Upsilon(1S)$ production in p--Pb collisions is suppressed at
forward rapidity, while at backward rapidity is consistent with unity. Models
including the nuclear modification of the gluon PDF or a contribution from
coherent parton energy loss tend to overestimate the measured $R_{\mathrm{pPb}}$
and cannot simultaneously describe the forward and backward rapidity
measurements.

\end{small}
\end{document}